\documentclass[final,3p,times,twocolumn]{elsarticle} 
\usepackage{graphicx}
\usepackage{epsfig} 
\usepackage{epstopdf}
\usepackage[skip=2pt,font=normalsize]{caption}
\usepackage{amssymb}
\usepackage{color}
\journal{Latex Template} 
\usepackage{lineno} 
\begin{document}  

\begin{frontmatter}


\title{BO-graphane and BO-diamane}



\author[1]{Babu Ram$^{\ast}$} 
\ead{babuaphy07@gmail.com}    
\address[1]{IBS-CMCM, Ulsan National Institute of Science and Technology (UNIST), Ulsan 44919, Republic of Korea} 
\author[2]{Rohit Anand} 
\address[2]{Department of Chemistry, Ulsan National Institute of Science and Technology (UNIST), Ulsan, Republic of Korea}
\author[3]{Arun S. Nissimagoudar}
\address[3]{OLA Battery Innovation Center, OLA Cell Technologies, Bangalore, India-560100}  
\author[2]{Geunsik Lee$^{\dagger}$}
\ead{gslee@unist.uc.kr}  
\author[1,2,4,5]{Rodney S Ruoff} 
\address[4]{Department of Materials Science and Engineering, Ulsan National Institute of Science and Technology (UNIST), Ulsan, Republic of Korea}
\address[5]{School of Energy and Chemical Engineering, Ulsan National Institute of Science and Technology (UNIST), Ulsan, Republic of Korea}

\begin{abstract}
The adsorption of boron and oxygen atoms onto mono- and multi-layer graphene leads to the formation of a buckled
graphene layer (BO-graphane) and a 2D diamond-like structure (BO-diamane) sandwiched between boron monoxide
layers per DFT calculations. BO-graphane has a calculated Young’s modulus ($E$) of 750 GPA and BO-diamane 771
GPa, higher than the calculated $E$ of -F, -OH, and -H diamanes; this is due to the presence of B-O bonds in the
functionalizing layers. Electronic band structure calculations show BO-graphane and BO-diamane are wide band gap
semiconductors with an indirect band gap up to a thickness of three layers (3L). Phonon dispersion and $ab-initio$
molecular dynamics (AIMD) simulations confirm dynamic and thermal stability, maintaining structural integrity at
1000 K. The room-temperature lattice thermal conductivity of BO-graphane and BO-diamane is found to be 879 W/m·K
and 1260 W/m·K, respectively, surpassing BeO (385 W/m·K), MgO (64 W/m·K), and Al$_{2}$O$_{3}$ (36 W/m·K); and F-diamane (377 W/m·K), and comparable to H-diamane (1145-1960 W/m·K), suggesting them as candidates for thermal
management in applications.
\end{abstract}
\end{frontmatter} 

\section{Introduction}
Two-dimensional carbon materials have been stud- ied extensively due to their properties and potential applications~\cite{RH1987,Bun1994,Cres1996,Narita1998}. Transformation from one type of carbon structure into another by ‘phase
engineering’~\cite{Ma2003,Song2013,babu2014,Pha2015,Yin2015}, opens possibilities for mechanical, electronic, optical, and thermal applications~\cite{ram2018,ramc568,Lavi2022}. Graphene consists of a single layer of carbon atoms and is a semimetal with a zero-band gap. Modifications in the electronic structure induced by stacking affect the functionality of graphene-based devices~\cite{Castro2009}. 
The electronic properties can also be tailored via chemical functionalization. In few-layer systems, surface chemistry, interlayer interactions, defect, and strain distribution become crucial, enabling a range of exotic properties combined with high flexibility and strength~\cite{Lavi2022}. The surface chemistry has a profound impact on the thermodynamic behavior of 2D materials, facilitating atomic rearrangement, modulating interface interactions, and enabling chemical bonding via surface functionalization. As an example, graphene, a semimetal, can be converted into graphane (C$_1$H$_{1}$), without external pressure or thermal energy. Graphane was first proposed theoretically in 2007~\cite{Sofo2007} but C$_1$H$_{1}$ has not been synthesized. In contrast, the fluorinated analog C$_1$F$_{1}$ that might be referred to as “F-graphane” has likely been synthesized. Theoretical studies show that the interaction of atomic hydrogen with a single layer of graphene induces corrugation, which consequently remove the conductive $\pi$ bands and results in the opening of a bandgap of up to 4.0 eV. However, the corrugation effect resulting from the sp$^{3}$ hybridization reduces the in-plane stiffness of graphane to 243 J/m$^{2}$, compared to 335 J/m$^{2}$ for pristine graphene~\cite{Ciraci2010}. To achieve and maintain the stability of diamane nanosheets under working conditions, the outer surfaces of graphene films can, in principle, be functionalized by hydrogen atoms~\cite{Elias2009}. The first theoretical investigation into converting bilayer graphene into a two-dimensional diamond structure, or H-diamane, through chemical functionalization is presented in~\cite{Cherno2009}. \\ 
Bakharev et al. described synthesis of F-diamane under ambient pressure conditions~\cite{Bak2020}. Studies have shown that graphene fluorination is an endothermic process that can be stabilized at 400 $^\circ$C, even in air, making the synthesis of fully fluorinated diamane (C$_2$F stoichiometry) possible while C$_2$H has not been made, just as C$_1$H$_{1}$ has not been been made but F-graphane has~\cite{Kva2017,clark2013,Nair2010}. Here, we used first-principles simulation that show thermodynamically stable materials through complete functionalization of monolayer and multilayer graphene with boron and oxygen atoms, referred to as BO-graphane and BO-diamane, respectively. We first explore the formation mechanisms of graphane and diamane nanosheets and evaluate their thermodynamic stability by calculating their formation energies relative to graphite. We investigated the dynamic, mechanical, thermal, and electronic properties of BO-graphane and BO-diamane. Hereafter, we refer to BO-graphane as 1L, while BO-diamanes with bilayer stacking (AB and AA) and a three-layer ABA-stacked configuration are denoted as 2L-AB, 2L-AA, and 3L, respectively. Additionally, we consider a single-side functionalized graphene sheet, referred to as ss-1L. We calculate the thermal conductivity of selected low energy configurations, specifically 1L and 2L-AB, using the full iterative solution of the Boltzmann transport equation as implemented in ShengBTE~\cite{Wu2014}.   
\\
\section{Methods}  
\begin{figure*}[!ht]
\centering
\includegraphics[width=\textwidth]{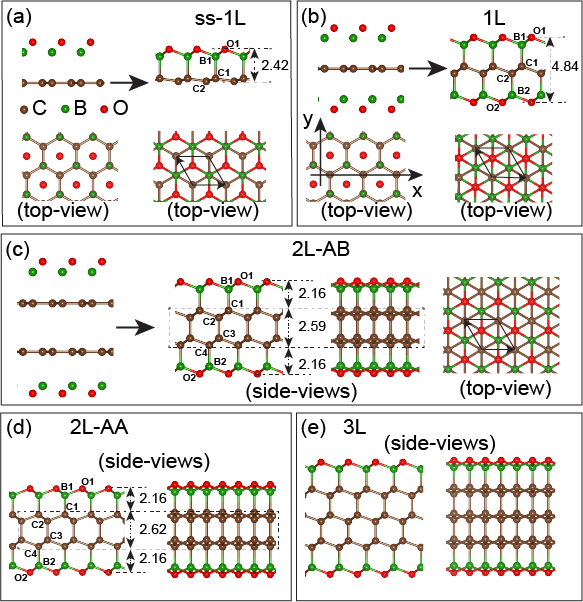} 
\caption{The mechanism of adsorption of boron and oxygen atoms (a) on a single-sided surface of a graphene sheet, (b) on both sides of graphene, and (c-d) in AB and AA stacked bilayer graphene. This leads to the formation of ss-1L, a buckled graphene layer (1L), as well as 2D diamond-like units (2L-AB and 2L-AA) sandwiched between a boron monoxide layer, with both side and top views provided. (e) shows a three-layer thick diamane configuration. The intra-layer thicknesses of each configuration are indicated within dotted lines, and the unit cells for each configuration are marked with black solid lines. The black arrows along the $x$ and 
$y$ directions indicate where strain is applied to assess mechanical properties. Carbon, boron, and oxygen atoms are represented by brown, green, and red solid spheres, respectively.}
\label{Fig:1} 
\end{figure*} 
All calculations were performed using first-principles density functional theory (DFT)~\cite{Kress1996, KressPRB1996}, as implemented in the Vienna Ab Initio Simulation Package (VASP)~\cite{Kress1996, KressPRB1996}. The projector augmented wave (PAW) method~\cite{Blochl1994, KresseJ1999} was used to represent electron-ion interactions. A plane-wave cutoff energy of 500 eV and a convergence criterion of 10$^{-5}$ eV were employed for the self-consistent electronic loop. The unit cells for the 1L, 2L (AA and AB), and 3L nanosheets consisted of 6, 8, and 10 atoms, respectively, with at least 15 \AA\ of vacuum added along the z-axis to avoid interactions between the sheet and its periodic images. A well-converged Monkhorst-Pack (MP) $k$-point mesh~\cite{Monkhorst1976} of $21\times21\times1$ was used for integrations over the Brillouin zone. The PHONOPY package was employed to acquire phonon dispersion and second-order force constants using density functional perturbation theory (DFPT)~\cite{togo2015}. The effect of van der Waals (vdW) interactions were included  using Grimme$^{'}$s DFT-D3 method~\cite{Grim2006}. \\ 
To examine the thermal stability of the nanosheets, we performed $ab-initio$ molecular dynamics (AIMD) simulations, using a time step of 1 femtosecond up to 45 ps in a $3\times3\times1$ supercell at a temperature of 1000 K. The lattice thermal conductivity and complex phononic properties were then calculated by iteratively solving the Boltzmann transport equation using the ShengBTE~\cite{Wu2014}. Third-order force constants were calculated with the thirdorder.py script~\cite{Wu2014}. For the third-order interatomic force constants (IFCs), a $4\times4\times1$ supercell, the same $k$-mesh, and energy cutoff were applied, with interactions considered up to the third nearest neighbors.
\\
\section{Results and discussion} 

\begin{table*}
\centering 
\caption{The optimized lattice parameter (a), bond distance, bond angle, and layer thickness ($\Delta(Z)$)  of ss-1L, 1L, 2L-AB, and 2L-AA nanosheets} 
\label{Table:1}
\begin{tabular}{c c c c c c} \\ 
\hline 
Structure (\AA) & $a$ (\AA) & Bond-distance (\AA) &  Bond-distance (\AA)   &   Angle ($^{\circ}$) & $\Delta(Z)$ (\AA)  \\          
\hline 
ss-1L & 2.51  & B1-O1=1.55      &   C1-C2=1.46   &  $\angle$B1-C1-C2=98.09$^\circ$ & 0.21 \\
      &       & B1-C1=1.67      &                      &  \\
      &       &                       &                      &   \\
1L    &  2.49 & B1(2)-O1(2)=1.54 &   C1-C2=1.53   &  $\angle$B1(2)-C1(2)-C2(1)=110.06$^{\circ}$  & 0.52 \\
      &       & B1(2)-C1(2)=1.60 &                      &    \\ 
      &       &                        &                      &  \\
2L-AB & 2.50 & B1(2)-O1(2)=1.54   &  C1(3)-C2(4)=1.53  & $\angle$B1(2)-C1(4)-C2(1)=109.83$^{\circ}$  & 2.59 \\  
      &      & B1(2)-C1(4)=1.60   &  C2-C3=1.55        &  $\angle$C1(2)-C2(3)-C3(4)=109.83$^{\circ}$       &        \\
      &      &                          &                          &                        \\      
2L-AA & 2.49 & B1(2)-O1(2)=1.54   & C1(3)-C2(4)=1.53   & $\angle$B1(2)-C1(4)-C2(1)=109.94$^{\circ}$ & 2.62 \\  
    &        & B1(2)-C1(4)=1.60   & C2-C3=1.57        & $\angle$C1(2)-C2(3)-C3(4)=109.94$^{\circ}$   &        \\
\hline \\  
\end{tabular} 
\end{table*}
The complete functionalization of monolayer by the boron and oxygen atoms on a single surface and on each surface leads to the formation of a hexagonal network of $sp^{2}$+$sp^{3}$ and $sp^{3}$-hybridized carbon atoms named as ss-1L, and 1L and shown in Fig.~\ref{Fig:1}(a-b). After optimization, ‘every other’ carbon atom of the graphene lattice moves off the plane and bond covalently with boron atoms, while boron atoms bond with oxygen atoms, forming a new 2D nanosheet made of carbon, boron, and oxygen. In bilayer graphene (AA and AB stacking), one carbon atom from one of the sublattices attaches to the functionalizing atoms, while the other (neighboring) carbon atoms bond with the layer below, creating a 2D diamond-like layer sandwiched between the boron and oxygen layers (Fig.~\ref{Fig:1}(c-d)). The structural parameters of ss-1L, 1L, 2L-AB, and 2L-AA are outlined in Table 1. The number of atoms in a unit cell for each configuration are ss-1L (C:2; B:1; O:1), 1L (C:2; B:2; O:2), 2L-AB (C:4; B:2; O:2), 2L-AA (C:4; B:2; O:2), and 3L (C:6; B:2; O:2), respectively. 
Similar to graphane~\cite{Topsa2010}, the optimized lattice constants and C-C bond distances for ss-1L (a = b = 2.51~\AA; C1-C2=1.46~\AA), 1L (a = b = 2.49~\AA; C1-C2=1.53~\AA), 2L-AB (a = b = 2.50~\AA; C1(2)-C2(3)=1.53-1.55~\AA), and 2L-AA (a = b = 2.49~\AA; C1(2)-C2(3)=1.53-1.57~\AA) (see, Table-1) increases compared to the graphene (a = b = 2.47~\AA; C1-C2=1.42~\AA)~\cite{Topsa2010}.
The observed C-C bond lengths (see, Table-1), bond angle ($\angle$ C1-C2-C3 = 109.83$^{\circ}$; 109.94$^{\circ}$), and average carbon-carbon interlayer distance of 2L-AB (2.065~\AA) and 2L-AA (2.093~\AA) closely match the experimental values of cubic diamond, which are 1.54~\AA, 109.5$^{\circ}$, and 2.05~\AA~\cite{Bak2020}. 
This supports the formation of a two-dimensional diamond layer (see supplementary information S1 (a-b)). When functionalizing the top and bottom surfaces of 3L, the carbon atoms in the outer graphene layers bond with the boron-oxygen layer, while the middle layer remains unaffected. \\
Achieving a thicker diamond film may require different pressure and temperature, as shown in Fig.~\ref{Fig:1}(e). The lattice parameters for 3L are a = b = 2.502~\AA\, with a carbon layer thickness of 4.65~\AA\ (Fig.~\ref{Fig:1}(e-f)). There is no significant change in the boron-carbon and boron-oxygen bond distances or lattice parameters for 3L compared to ss-1L, 1L-AB, 2L-AB, and 2L-AA, indicating that these properties are not sensitive to the thickness of the carbon layer. 

\subsection{Formation energy}

\begin{figure*}[!ht]
\centering
\includegraphics[width=0.95\textwidth]{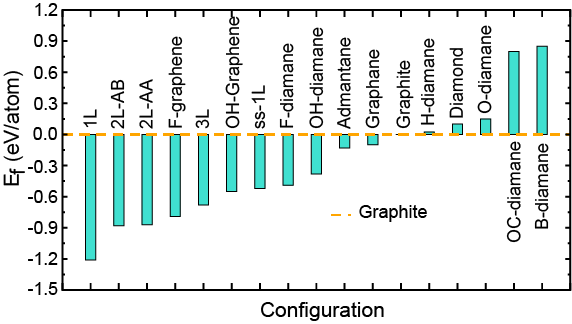} 
\caption{Formation energy of theoretically predicted (1L, 2L-AB, 2L-AA, 3L, ss-1L, O-diamane, OC-diamane, and B-diamane) and experimentally synthesized (F-graphene, Graphane, F- and H-diamane, OH-diamane) nanosheets. Formation energy is taken with respect to graphite.}
\label{Fig:2} 
\end{figure*}

\begin{figure}[!ht]
\centering
\includegraphics[width=0.95\columnwidth]{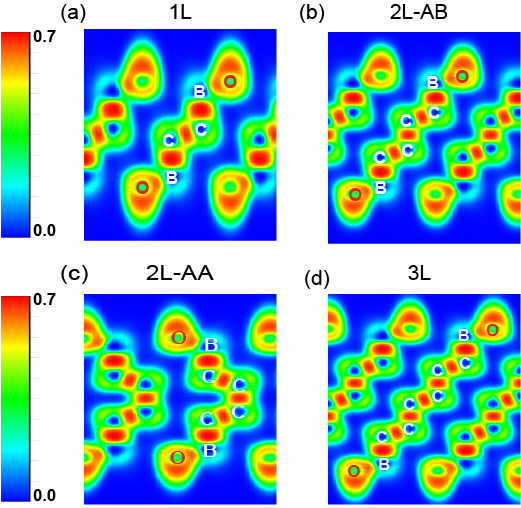} 
\caption{The electron localization function (ELF) for (a) 1L, (b) 2L-AB, (c) 2L-AA, and (d) 3L nanosheets. In this representation, red indicates localized charge density regions, while light-green represents delocalized charge density, with the color bar providing a reference.}
\label{Fig:3} 
\end{figure}  

We investigated the thermodynamic, dynamical, thermal, and mechanical stability of the reported nanosheets. To assess thermodynamic stability, we
calculate the formation energy (E$_f$ ) as follows: 

\begin{equation}
E_f=\frac{E_{CBO}-{n_C}E_{C}-{n_B}E_{B}-{n_O}E_{O}}{n_C+n_B+n_O} 
\end{equation}
In this equation, $E_{CBO}$, $E_{C}$ (-9.31 eV), $E_{B}$ (-6.81 eV), $E_{O}$ (-3.39 eV) represent the total energy of the structure,
the energy of a carbon atom in graphite, energy per atom of $\alpha$-boron, and the energy of an oxygen atom in
an O$_{2}$ molecule. Here, $n_{C}$, $n_{B}$, and $n_{O}$ are the number of carbon, boron, and oxygen atoms in a unit cell for each configuration. Among single-layer functionalized 2D materials such as graphane, F-graphene, OH-graphene, and their diamane counterparts (H-diamane, F-diamane, and OH-diamane), the formation energies of 1L and 2L- AB are the lowest, as shown in Fig.~\ref{Fig:2}. When considering stacking types, Bernal AB-stacking is the most energetically favorable, even after hydrogenation or fluorination. The 2L-AB stacked diamane has a lower formation energy than the 2L-AA, but the difference is only 0.01 eV/atom (Fig.~\ref{Fig:2}). This makes both 2L-AB and 2L-AA configurations good candidates for experimental realization, regardless of the stacking type. The formation energy difference between 1L and existing functionalized sheets (F-graphene, H-graphene, and OH-graphene) is 0.42 eV/atom, 1.1 eV/atom, and 0.66 eV/atom, respectively (Fig.~\ref{Fig:2}). This significant difference in formation energy compared to well-known single-layer functionalized graphene indicates that 1L has higher thermodynamic stability. 2L-AB has a formation energy that is lower than F-diamane (by 0.39 eV/atom), H-diamane (by 0.86 eV/atom), OH-diamane (by 0.50 eV/atom), and adamantane (by 0.75 eV/atom). The O-diamane, B-diamane, and OC-diamane (where B-atoms are replaced by C-atoms) have higher formation energies than diamond (Fig.~\ref{Fig:2}) and were thus not considered for further study here. \\
Although the formation energy of the 3L thick layer is higher than that of the mono- and bilayers (Fig.~\ref{Fig:1}(e-f)), it is still lower than that of the F-diamane, H-diamane, and OH-diamane. The formation energy of ss-1L is higher than that of  1L, 2L-AB, 2L-AA, and 3L, but comparable to or lower than X-diamane (X=F, H, OH). We note that ss-1L does not maintain structural integrity even at 300 K (see supplementary information S2 (a)), thus was also not studied further. From Fig.~\ref{Fig:3}(a-d), it is evident that electrons are more localized between B-C and C-C atoms, indicating the formation of strong covalent bonds between these atoms, with maximum iso-surface values of 0.7. The charge distribution near B-atoms is weak, while it is strongly localized around oxygen atoms due to its higher electronegativity (3.44) compared to boron (2.04), which represents the polar covalent character of the B-O bond. Additionally, the electronegativity difference is greater between B-O than between C-B. Per Bader analysis, 1.5 e to 1.58 e is transferred from boron to oxygen and 0.50 e to 0.55 e from boron to carbon atoms (immediately bonded to boron) across 1L, 2L-AB, 2L-AA, and 3L. Furthermore, charge localization is relatively greater between B-C atoms than C-C atoms, indicating stronger interactions between B-C atoms. We can expect this effect to contribute to the increase in formation energy, as the influence of the boron-oxygen layers diminishes with the thickness of the carbon layer
\\
\subsection{Dynamic stability and thermal conductivity} 

\begin{figure}[!ht]
\centering
\includegraphics[width=0.95\columnwidth]{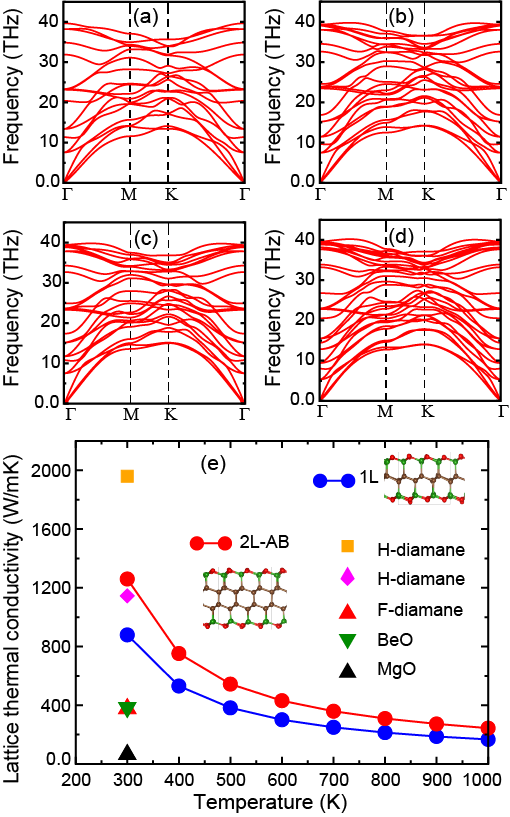} 
\caption{The phonon dispersion curves for (a) 1L, (b) 2L-AB, (c) 2L-AA, and (d) 3L are presented. (e) Shows the lattice thermal conductivity as a function of temperature for the selected low-energy configurations, 1L (blue curve) and 2L-AB (red curve), up to 1000 K. For comparison, the thermal conductivities of H-diamane~\cite{MoRR2020,Zhu2019}, F-diamane~\cite{MoRR2020}, BeO, and MgO~\cite{Mor2021} are also included.} 
\label{Fig:4} 
\end{figure}
The phonon dispersion along high-symmetry directions at 0 K was calculated. The absence of imaginary frequencies indicates that 1L, 2L-AB, 2L-AA, and 3L are kinetically stable, as shown in Fig.~\ref{Fig:4}(a-d). Compared to F-diamane (see supplementary information S2 (c)), the low-frequency acoustic phonon modes in the phonon spectrum of 2L-AB (Fig.~\ref{Fig:4}(b)) are more hard, reflecting greater dynamic stability. This greater stability may also explain why 2L-AB has a lower formation energy than F-diamane. We explored the lattice thermal conductivity of the more energetically stable configurations, 1L and 2L-AB. Using the full iterative solution of the Boltzmann transport equation with ShengBTE~\cite{Wu2014}, we found that the room temperature lattice thermal conductivity of 1L is 879 W/m.K, and that of 2L-AB is 1260 W/m.K. For comparison: BeO (385 W/m.K), MgO (64 W/m.K), and Al$_{2}$O$_{3}$ (36 W/m.K)~\cite{Mor2021,Xi2010}, as shown in Fig.~\ref{Fig:4}(e). And: h-BN (545 W/m·K)~\cite{Ying2020}, MoS$_{2}$ (84 W/m·K)~\cite{Zhang2015}, and MoSe$_{2}$ (59 W/m.K)~\cite{Zhang2015}. The thermal conductivity of 1L and 2L-AB decreases with increasing temperature, and is calculated as 167 W/m.K and 244 W/m K at 1000 K (Fig.~\ref{Fig:4}(e)), respectively. Notably, 2L-AB has significantly higher calculated thermal conductivity than F-diamane (377 W/m.K)~\cite{MoRR2020} and is comparable to H-diamane (1145-1960 W/m.K)~\cite{MoRR2020,Zhu2019}. Generally, thermal conductivity is controlled by two mechanisms. First, the wide or narrow dispersion of each mode, particularly the acoustic modes (LA, TA, ZA), significantly affects the group velocity and, consequently, the thermal conductivity. Here, LA modes are associated with in-plane vibrations, while TA modes involve in-plane vibrations perpendicular to the direction of propagation. \\
Flexural out-of-plane acoustic (ZA) branches exhibit quadratic behavior near the zone center, particularly in 2D materials. Second, the maximum frequency of the acoustic modes and the point where the acoustic and lowest optical mode (LOM) branches intersect also influence thermal conductivity. The acoustic branches of 1L and 2L-AB are wider than those of F-diamane (S2(b-c)). The latter has a lower maximum frequency for its acoustic branches, significantly below 8 THz, corresponding to the LA phonon mode (S2(b-c)). Additionally, F-diamane has narrow dispersion branches that intersect with the lower optical modes, leading to increased phonon scattering and reduced thermal conductivity. In contrast, the LA and TA branches of 1L and 2L-AB are more dispersive from the zone center, with maximum frequencies around 12 THz. The TA phonon branch of 2L-AB is steeper than that of 1L, resulting in a higher group velocity and increased thermal conductivity (Fig.~\ref{Fig:4}(a-b) and S2(b-c)). Similarly, the acoustic branches in H-diamane are wider and do not interfere with other modes, leading to higher or comparable thermal conductivity compared to other diamane types~\cite{MoRR2020} as shown in Fig.~\ref{Fig:4}(e). To gain more insight into the thermal response of 1L and 2L-AB nanosheets, we obtained the anharmonic scattering rate at 300 K (Figure S2(d)). This analysis helped us explore the microscopic origins of the differences in thermal conductivity. Figure S2(d) shows the phonon scattering rate as a function of phonon frequency, highlighting the degree of anharmonicity. Notably, in both the low and high-frequency phonon ranges, 1L exhibits significantly higher scattering rates compared to 2L-AB. This increased phonon scattering leads to reduced thermal conductivity in 1L compared to 2L-AB.
\\
\subsection{Mechanical stability} 

\begin{figure*}[!ht]
\centering
\includegraphics[width=0.95\textwidth]{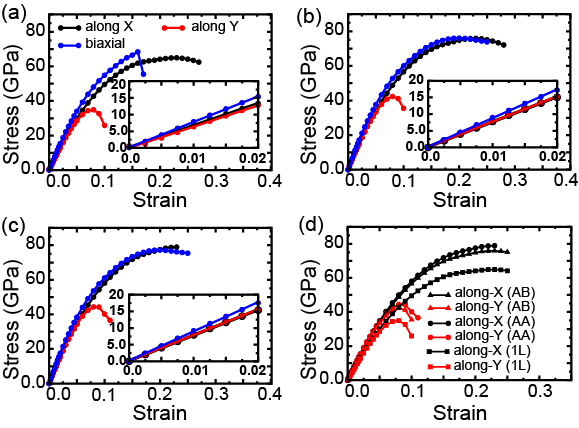} 
\caption{The stress-strain response along uniaxial and biaxial directions for (a) 1L, (b) 2L-AB, and (c) 2L-AA is shown, along with a comparison of the stress-strain responses for the 1L, 2L-AB, and 2L-AA configurations (d). The inset displays the linear regression of the stress-strain curves within 0.02 of applied strain to determine the elastic moduli.}
\label{Fig:5} 
\end{figure*}

\begin{table}
\small  
\caption{The predicted elastic moduli ($E$), maximum stress corresponding to the applied strain ($\sigma^{c}$), critical strain ($\epsilon^{c}$), dynamic stability (DS), and thermal stability (TS) at 1000 K for 1L, 2L-AB, 2L-AA, 3L, F-AB, and OH-AB nanosheets in the $x$, $y$, and biaxial directions.} 
\label{Table:1}
\begin{tabular}{c c c c c c} \\ 
\hline 
Layer & ($E$) & $\sigma^{c}$ & $\epsilon^{c}$ & DS  & TS \\          
      &  (GPa)     &  (GPa)       &   ($\%$)       & 0 K  & (1000 K) \\
\hline
\hline 
(1L)    &         &      &       &  yes   &  yes    \\ 
${x}$   & 659.66  & 64.91 & 0.23 & &  \\ 
${y}$   & 670.58  & 34.89 & 0.08 &  &  \\
biaxial & 765.83  & 68.41 & 0.16 &  &  \\

\hline
(2L-AB) &         &      &       &  yes   &  yes    \\ 
${x}$   & 750.63  & 75.90 & 0.23 &  &  \\ 
${y}$   & 755.58  & 42.35 & 0.08  &  &  \\
biaxial & 859.17  & 75.87 & 0.21  &  &  \\

\hline
(2L-AA) &         &        &      &  yes &  yes   \\ 
${x}$   & 764.98  & 78.94  & 0.23 &  & \\ 
${y}$   & 771.13  & 44.35  & 0.08 &  & \\
biaxial & 875.41  & 77.18  & 0.20 &  &  \\

\hline
(3L)          &        &      &       &  yes   &  yes   \\     
${x}$   & 804.42 & 82.90 & 0.23 &  &  \\ 
${y}$   & 812.33 & 47.02 & 0.08 &  &  \\
biaxial & 907.41 & 78.97 & 0.19 &  &  \\

\hline
(F-AB)          &        &      &       &  yes~\cite{MORT2020}   &  yes~\cite{MORT2020}   \\     
${x}$   & 542.66  & 56.18 & 0.23  &  &  \\ 
${y}$   & 547.69  & 44.39 & 0.14  &  &  \\
biaxial & 584.73  & 44.15 & 0.17  &  &  \\

\hline
(OH-AB)          &        &      &       &     &    \\ 
${x}$   & 479.64  & 47.83 & 0.20 &  &  \\ 
${y}$   & 470.36  & 33.77 & 0.12 &  & \\
biaxial & 550.35  & 39.67 & 0.15 &  &  \\
\hline \\ 
\end{tabular} 
\end{table}

\begin{figure*}[!ht]
\includegraphics[width=0.95\textwidth]{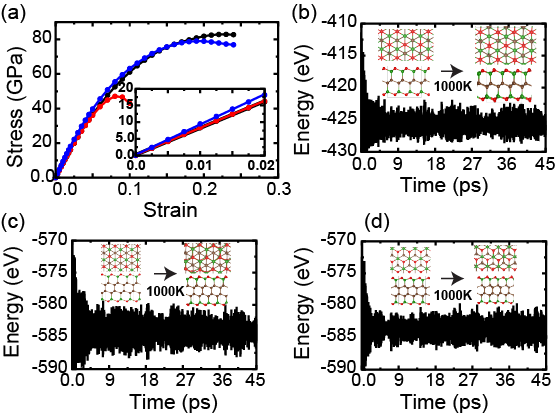} 
\caption{(a) The stress-strain response of 3L along uniaxial and biaxial directions is analyzed to obtain the elastic moduli within 0.02 of applied strain (inset). The inset shows the linear regression of the stress-strain curves within this range. (b-d) The total energy as a function of simulation time using AIMD, with a time step of 1 fs up to 45 ps at 1000 K. The initial and final configurations (top/side views) are shown in the inset, with the final configuration taken after 45 ps of simulation time.}
\label{Fig:6}  
\end{figure*}
The mechanical properties of these nanosheets was studied by applying uniaxial and biaxial tensile strain. A rectangular unit cell is used to calculate the stress-strain response. Because there is a vacuum in the direction normal to the sheet, the stress in that direction remains negligible after loading followed by the geometric optimization. The geometric optimization accounts for the Poisson’s ratio effect, which usually causes stress perpendicular to the loading direction. For calculating the stress-strain response, the thickness of the sheet is defined as the distance between the outermost atoms on each side, plus their respective van der Waals radii. Additionally, due to the vacuum in the normal direction, we needed to rescale the in-plane stress component of the nanosheets. The calculated stress-strain curves show an initial linear region followed by a nonlinear region based on linear elasticity. We identified a harmonic region within 0.02 of the applied strain and used linear regression (inset) to calculate the 2D elastic modulus ($E$), as shown in inset Fig.~\ref{Fig:5}(a-d), Fig.~\ref{Fig:6}(a), and supplementary information S2 (e-f).  By fitting the initial slope of the stress-strain curve with linear regression, we obtained the elastic moduli along-$x$ ($E_x$), $y$ ($E_y$), and biaxial ($E_{biaxial}$) directions, as given in Table-2. The elastic modulus is a measure of bond strength; therefore, a larger elastic modulus indicates stiffer bonds. The elastic moduli in the two orthogonal directions do not vary significantly, indicating that the new configurations, along with F-diamane, H-diamane~\cite{Morz2020}, and OH-diamane, are elastically isotropic (see Table-2). Our calculated elastic moduli $E_x$ and $E_y$ for F-diamane also agree with those reported by Mortazavi et al~\cite{Morz2020}. The maximum stress corresponding to the applied strain for 2L-AB and 2L-AA is very similar and higher than that of 1L, suggesting that the stacking method has minimal impact on the ultimate tensile strength (UTS) and corresponding fracture strain, as shown in Fig.~\ref{Fig:5}(d). A similar trend in stress-strain behavior for AA and AB stacking types can also be observed in H-diamane~\cite{Zhu2024}. \\
Our predicted elastic moduli for 2L-AB, 2L-AA, and 3L are much larger than those of F-diamane and OH- diamane of similar thickness, while being comparable to H-diamane~\cite{Morz2020} (see, Table-2). The elastic modulus ($E$) and critical strain ($\sigma^{c}$), which is the strain at maximum tensile strength, increase with carbon layer thickness, as expected when approaching the bulk diamond limit. The influence of the functional group decreases as thickness increases. Our results indicate that the strain corresponding to the maximum strain limit ($\epsilon^{c}$) remains constant regardless of layer thickness and stacking type (see, Table-2). As the applied strain increases further, the stress-strain curve diverges in the two orthogonal directions. Specifically, the maximum stress in the $x$-direction is higher than in the $y$-direction, indicating anisotropy in the stress-strain response and demonstrating a greater load-bearing capacity compared to known F-diamane, H-diamane~\cite{Morz2020}, and OH-diamane (see, Table-2). This anisotropic behavior in response to applied strain is also observed in existing F-diamane and H-diamane~\cite{Morz2020} and OH-diamane. With a goal of understanding the anisotropy in the stress-strain response, we examined the effects of strain on bond length and bond angle between carbon atoms, as shown in supplementary information S3 (a-f). When applying uniaxial strain, the bonds aligned with the strain direction stretch more, while those in the perpendicular direction change slightly due to the Poisson effect. In the case of biaxial strain, bond length and angle changes uniformly until the structure reaches a limit where it begins to break. In bilayers (2L-AB and 2L-AA), similar to the 1L case, the C-C bonds along the strain direction respond more than those in the perpendicular direction. However, because of the increased thickness of the diamanes, bond lengths in the off-plane direction decrease due to the compression effect caused by in-plane loading, as shown in supplementary information S4 (a-f) and S5 (a-f). Similar effects can also be observed in the variation of bond angles with the direction of applied strain in the 1L, 2L-AB, and 2L-AA cases (see supplementary information S3/S4/S5 (a-f)). These 2D materials, exhibiting high mechanical strength, could be particularly attractive for ultra-thin protective coatings and as ultra-high-strength components in composite materials~\cite{Piz2019}. \\
\subsection{Thermal stability} 

We used ${ab-initio}$ molecular dynamics (AIMD) to assess thermal stability. In AIMD simulations, we estimated the total energy variation over 45 ps with a time step of 1 fs at a temperature of 1000 K, as shown in Fig.~\ref{Fig:6}(b-d). The ss-1L configuration cannot maintain its structural integrity even at 300 K, leading to a drastic change in total energy over time (see supplementary information S2(a)). This indicates the need for complete coverage of boron and oxygen on both sides to stabilize the system. We found that the final configuration after 45 ps does not change significantly, keeping the 1L, 2L-AB, 2L-AA (Fig.~\ref{Fig:6}(b-d) (inset)) and 3L intact (inset, supplementary information S6(a)). Additionally, the total energy variation remains nearly unchanged over the simulation time. Further investiga- tion of structural resistance at 1000 K involved plotting the average bond lengths over time for 1L, 2L-AB, 2L- AA, and 3L, respectively, as shown in supplementary information S6(b-e)). There is no significant increase in the B-O and B-C bond lengths as we move from 1L to 3L. However, a slight increase in C-C bond length is observed with thickness, as the C-C bond length approaches its bulk diamond limit. \\
\subsection{Electronic properties}

\begin{figure*}[!ht] 
\includegraphics[width=\textwidth]{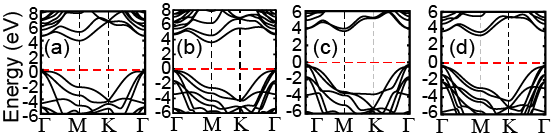} 
\caption{The electronic band structures for (a) 1L, (b) 2L-AB, (c) 2L-AA, and (d) 3L are shown. The red dotted line represents the Fermi level.}
\label{Fig:7} 
\end{figure*}

\begin{figure}[!ht]
\centering
\includegraphics[width=\columnwidth]{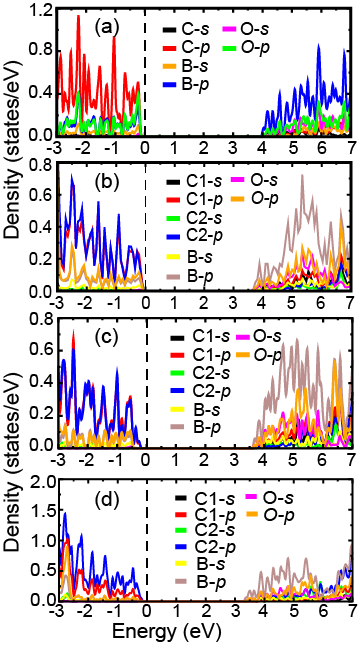} 
\caption{The density of states for (a) 1L, (b) 2L-AB, (c) 2L-AA, and (d) 3L nanosheets predicted using the PBE functional is shown. The black dotted line represents the Fermi level.}
\label{Fig:8} 
\end{figure}
We examined the electronic properties of these nanosheets by analyzing their electronic band structures and the angular momentum resolved density of states (LDOS) using the PBE functional. The boron-oxygen passivated nanosheets exhibit semiconducting behavior with a wide band gap. The 1L nanosheet has a band gap of 4.22 eV, with the valence band maximum (VBM) located between the $\Gamma$- and M points, and the conduction band minimum (CBM) at the M point, making it an indirect band gap semiconductor, as shown in Fig.~\ref{Fig:7}(a). Regardless of stacking type, the BO-diamanes (2L-AB and 2L-AA) also demonstrate indirect band gaps of 3.87 eV and 3.91 eV, with band edges at the $\Gamma$- and M points, as shown in Fig.~\ref{Fig:7}(b-c)). Even as we increase the thickness up to 3L, the band gap remains indirect, measuring 3.55 eV (Fig.~\ref{Fig:7}(d)). However, the band gap decreases with increasing layer thickness, a trend also observed in H-diamane~\cite{Morz2020}. This decrease could be due to the weakening of the quantum confinement effect perpendicular to the plane. In addition, the electronic band structure of a material is primarily determined by the contributions of atomic orbitals to the band edges. The contributions from different orbitals at the band-edge positions for 1L, 2L-AB, 2L-AA, and 3L are shown in Fig.~\ref{Fig:8}(a-d). For 1L, the VBM is mainly composed of the out-of-plane $\it{p_z}$ orbitals of the C, B, and O atoms, while the CBM is primarily made up of the in-plane B-$\it{p_x}$-orbitals and the out-of-plane $\it{p_z}$ orbitals of the oxygen atoms, as shown in S7(a- c). There is also a small contribution to the VBM from the in-plane C-$\it{p_x}$/${p_y}$ orbitals (S7(a-c)). \\
In the case of 2L-AB-diamane, the VBM receives contributions from the hybridization of off-plane $\it{p_z}$-orbitals of C1, C2, B, and O atoms, similar to the 1L case. The CBM primarily comes from B-$\it{p_x}$-orbitals, with smaller contributions from B-$\it{p_y}$/${p_z}$, C1-$\it{s}$, and O-$\it{s}$/${p_z}$ surface states (S8 (a-d)). In this context, C1 refers to the carbon atoms attached to the boron atoms, while C2 atoms are bonded only to carbon atoms (see S8(a-d)). Due to  the change in stacking order from AB to AA, the 2L-AA diamane shows dominance of the in-plane $\it{p_y}$/${p_y}$ orbitals of C1 and C2 atoms at the VBM, unlike the 2L-AB, where the VBM contribution comes from off-plane orbitals (see S9(a-d)). The antibonding states are primarily occupied by B-$\it{p_y}$/${p_y}$ and O-$\it{p_z}$/${s}$ orbitals, with a minor contribution from the C1-$\it{s}$-orbital (S9(a-d)). For the 3L configuration, the VBM and CBM are formed by the in-plane $\it{p_x}$/${p_y}$ orbitals of C1 and C2 and the B-$\it{p_x}$/${p_y}$/${p_z}$  orbitals (S10 (a-d)). As the thickness increases from 1L to 3L, the presence of surface states of oxygen atoms at the VBM and CBM decreases due to the quantum confinement effect. Overall, the combined effects of surface states and non-surface states, along with the quantum confinement effect, significantly influence the band-edge states and the tuning of the band gap as the thickness of the nanosheets changes. However, the indirect nature of the band gap remains unchanged and insensitive to the thickness of the nanosheets, which could be beneficial for applications in photocatalysis and nano-electronics. \\ 
\section{Summary}

We conducted a first-principles study to explore two-dimensional (2D) materials by adsorbing boron and oxygen atoms onto mono- and multilayer graphene surfaces, referred to as BO-graphane and BO-diamane. These nanosheets serve as a 2D counterpart to graphane, and the latter possesses a diamond-like structure. We examined their stability, mechanical properties, electronic properties, and thermal conductivity. Both BO-graphane and BO-diamane exhibit excellent thermodynamic and dynamic stability, along with high mechanical strength, and structural integrity even at 1000 K. This is supported by formation energy, phonon dispersions, and ab initio molecular dynamics (AIMD) simulations. Our analysis shows isotropic elasticity and a highly anisotropic tensile response. The elastic moduli and strength increase with the number of carbon layers, exceeding those of existing diamane nanosheets like H-diamane, F-diamane, and OH-diamane. Electronic band structure calculations indicate that these unique structures, consisting of carbon, boron, and oxygen atoms, are wide-gap semiconductors. BO-graphane and BO-diamane have higher calculated thermal conductivity than BeO, MgO, and Al$_{2}$O$_{3}$. They also show superior thermal conductivity compared to F-diamane and are comparable to H-diamane. This study presents a novel approach to constructing and stabilizing diamane structures and offers valuable insights into their properties. We hope that these findings will serve as a useful guide for future research. 


\section*{Acknowledgements} 

We appreciate the financial support from the Institute of Basic Science grant number (IBS-R019-D1). The DFT calculations were performed using the UNIST computational facilities and the supercomputer at the Korea Institute of Science and Technology Information (KISTI). 
\bibliographystyle{elsarticle-num-names} 
\bibliography{CBO.bib} 


\end{document}


\maketitle

\section*{Supplementary Figures}

\begin{figure*}[!ht] 
\centering
\small
\includegraphics[width=0.95\textwidth]{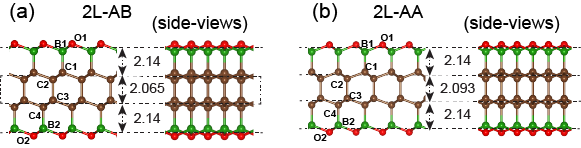} 
\caption{(a-b)2L-AB and 2L-AA stacked diamane are shown in side views, with an intra-layer distance of 2.14 \AA\ between carbon and boron monoxide. The average thickness of the carbon layer (or 2D diamond unit) sandwiched between the boron monoxide layers are 2.065 \AA\ and 2.093 \AA. In the representation, oxygen atoms are shown as red spheres, boron atoms as green spheres, and carbon atoms as brown spheres.} 
\label{Fig:S1} 
\end{figure*}

\begin{figure*}[!ht] 
\centering
\small
\includegraphics[width=0.7\textwidth]{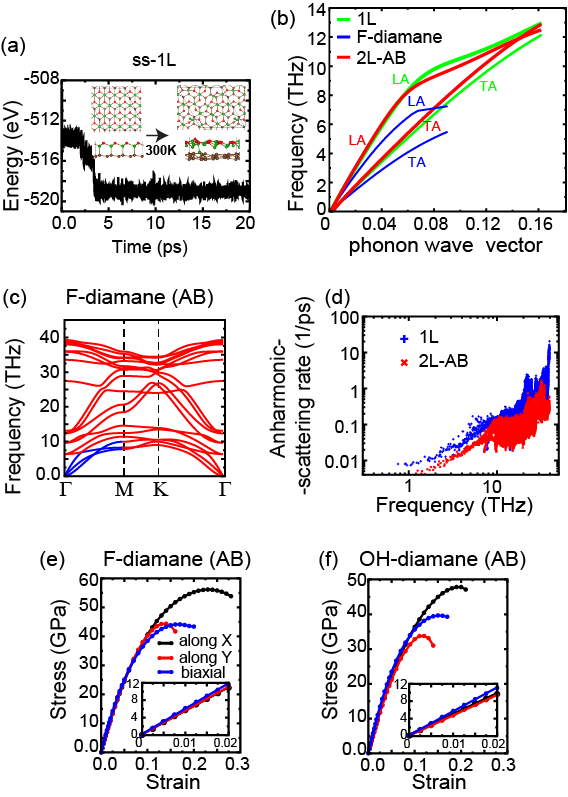} 
\caption{(a) Total energy variations as a function of simulation time for ss-1L sheet (top and side views) at the end of the 20 ps AIMD simulations at 300K , respectively. (b) Shows the frequencies of acoustic phonon branches (LA and TA) at the zone center as a function of phonon wave vector for 1L, 2L-AB and F-diamane, respectively. (c) phonon dispersion speatra of F-diamane. Blue colored bands represents accoustic phonon modes. (d) The anharmonic scattering rate as a function of phonon frequency is plotted on a logarithmic scale for 1L and 2L-AB at 300 K. (e-f) Stress-strain response curve of AB-stacked F-diamane and OH-diamane along $x$, $y$, and biaxial directions. Linear region (inset) of stress-strain curves are plotted within 0.02 of applied strain to calculate the elastic moduli $x$, $y$, and biaxial directions.}
\label{Fig:S2} 
\end{figure*}

\begin{figure*}[!ht]  
\centering
\small
\includegraphics[width=0.8\textwidth]{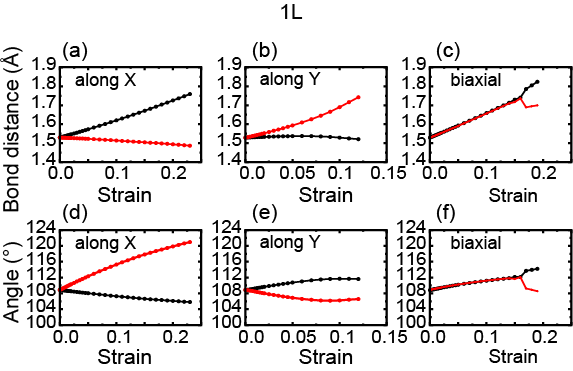} 
\caption{(a-f) The bond lengths and bond angles between carbon atoms as a function of applied strain in the $x$, $y$, and biaxial directions for 1L are shown. The x and y directions of applied strain are indicated in Fig. 1(b)} 
\label{Fig:S3} 
\end{figure*} 

\begin{figure*}[!ht] 
\centering
\small 
\includegraphics[width=0.95\textwidth]{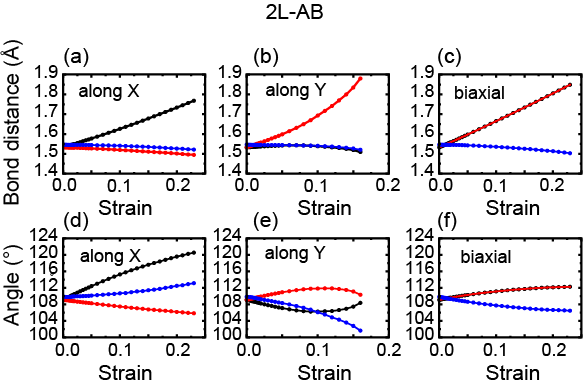} 
\caption{(a-f) We plotted the bond lengths and bond angles (between carbon atoms) as a function of applied strain in the $x$, $y$, and biaxial directions for 2L-AB. The directions of the applied strain are indicated in Fig. 1(b)}   
\label{Fig:S4} 
\end{figure*}

\begin{figure*}[!ht]  
\centering
\small
\includegraphics[width=0.95\textwidth]{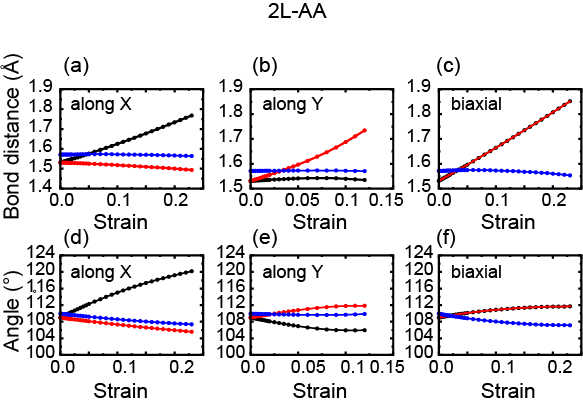} 
\caption{ (a-f) For the 2L-AA stacked diamane, we present the variation in bond lengths and bond angles (between carbon atoms) with applied strain in the $x$, $y$, and biaxial directions. The $x$, $y$ directions of applied strain are shown in Fig. 1(b)}
\end{figure*} 

\begin{figure}[!ht]
\centering
\includegraphics[width=0.65\textwidth]{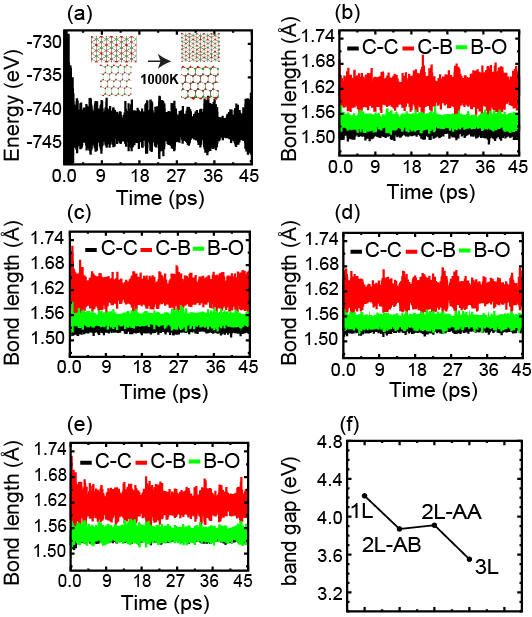}
\caption{(a) For the 3L case, the total energy is shown as a function of simulation time over 45 ps at 1000 K, with a time step of 1 fs. The initial and final atomic configurations at the end of the simulation are displayed in the inset (top and side views). (b-e) Show the average bond lengths between C-C (black solid line), C-B (red solid line), and B-O (green solid line) as a function of simulation time over 45 ps at 1000 K with a time step of 1 fs. (f) Illustrates the variation of the electronic band gap with the thickness of the carbon layers for 1L, 2L-AB, 2L-AA, and 3L.} 
\label{Fig:S6}
\end{figure}

\begin{figure}[!ht]  
\centering
\includegraphics[width=0.6\textwidth]{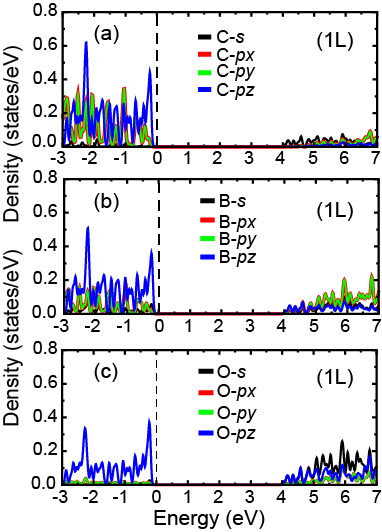} 
\caption{ (a-c) The angular momentum resolved density of states (LDOS) for 1L is shown. The black dotted line represents the Fermi energy level, while C, B, and O denote carbon, boron, and oxygen atoms, respectively.} 
\label{Fig:S7}
\end{figure} 

\begin{figure}[!ht]
\centering
\includegraphics[width=0.6\textwidth]{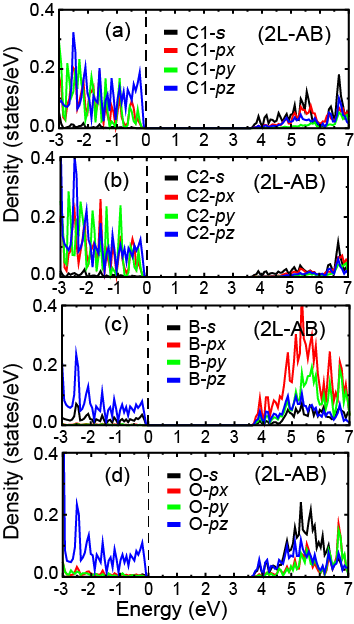}
\caption{(a-d) The angular momentum resolved density of states (LDOS) for 2L-AB is shown. The black dotted line represents the Fermi energy level, while C, B, and O indicate carbon, boron, and oxygen atoms, respectively.}
\label{Fig:S8}
\end{figure}

\begin{figure}[!ht]  
\centering
\includegraphics[width=0.6\textwidth]{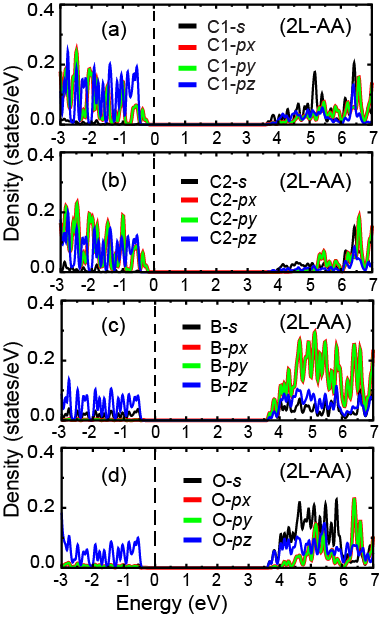} 
\caption{ (a-d) The angular momentum resolved density of states (LDOS) for 2L-AA is presented. The black dotted line represents the Fermi energy level, with C, B, and O denoting carbon, boron, and oxygen atoms, respectively.} 
\label{Fig:S9}
\end{figure} 

\begin{figure}[!ht]
\centering
\includegraphics[width=0.6\textwidth]{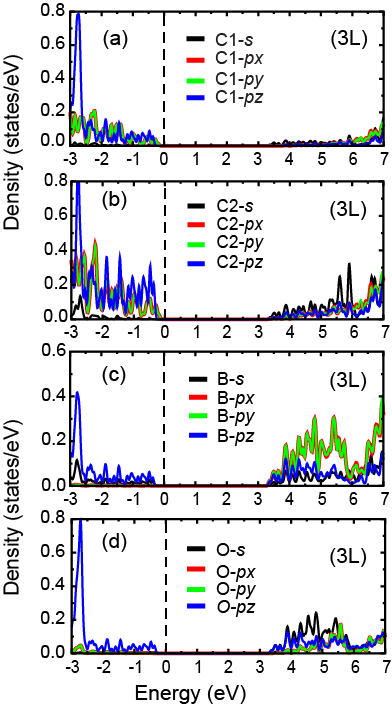}
\caption{(a-d) The angular momentum resolved density of states (LDOS) for 3L is shown. The black dotted line indicates the Fermi energy level, with C, B, and O representing carbon, boron, and oxygen atoms, respectively.}
\label{Fig:S10}
\end{figure}